# Machine-learning based interatomic potential for phonon transport in perfect crystalline Si and crystalline Si with vacancies

Hasan Babaei[1], Ruiqiang Guo[1], Amirreza Hashemi[2], Sangyeop Lee*[1,3]

[1]Department of Mechanical Engineering and Materials Science, University of Pittsburgh, Pittsburgh, Pennsylvania 15261, USA

[2]Department of Computational Modeling and Simulation, University of Pittsburgh, Pittsburgh, Pennsylvania 15261, USA

[3]Department of Physics and Astronomy, University of Pittsburgh, Pittsburgh, Pennsylvania 15261, USA

* sylee@pitt.edu

**Abstract**

We report that single interatomic potential, developed using Gaussian regression of density functional theory calculation data, has high accuracy and flexibility to describe phonon transport with *ab initio* accuracy in two different atomistic configurations: perfect crystalline Si and crystalline Si with vacancies. The high accuracy of second- and third-order force constants from the Gaussian approximation potential (GAP) are demonstrated with phonon dispersion, Grüneisen parameter, three-phonon scattering rate, phonon-vacancy scattering rate, and thermal conductivity, all of which are very close to the results from density functional theory calculation. We also show that the widely used empirical potentials (Stillinger-Weber and Tersoff) produce much larger errors compared to the GAP. The computational cost of GAP is higher than the two empirical potentials, but five orders of magnitude lower than the density functional theory calculation. Our work shows that GAP can provide a new opportunity for studying phonon transport in partially disordered crystalline phases with the high predictive power of *ab initio* calculation but at a feasible computational cost.

Modeling of materials at atomic scale requires an access to energies and forces on the atoms. These quantities are obtained using either *ab initio* calculations or empirical potentials (EPs). The *ab initio* calculations provide the highest accuracy, but are not feasible for large systems due to the high computational cost. The EPs have predefined functional forms with parameters that are optimized so that the model reproduces either some macroscopic physical properties (e.g., elastic constants) or forces from quantum calculations. Although the simulations using EPs are much less expensive than the *ab initio* calculations, they often show a poor accuracy when used to predict physical quantities that are not employed to optimize the parameters. Also, because of the rigid functional forms, EPs are usually not transferrable to other phases.

Among many physical quantities, thermal conductivity is one quantity that both EPs and direct *ab initio* calculation have severe limits for the prediction. The thermal transport in crystalline materials is often described with quasi-harmonic phonon gas that have subtle anharmonic interaction. For crystalline Si, as an example, the EPs fail to accurately reproduce the phonon dispersion [1] and the thermal conductivity values [2]; the thermal conductivity values at 300 K from the two most widely used EPs, the Stillinger-Weber (SW) [3] and Tersoff [4,5] potentials, are ~700 W/m-K and ~350 W/m-K [2], respectively, largely deviating from the experimental value of ~150 W/m-K [6]. The *ab initio* Boltzmann transport method, which employs the phonon gas model, successfully predicts the thermal conductivity values of various crystalline materials [7,8]. However, the phonon gas model fails in partially or fully disordered phase where the vibrational eigenmodes are non-propagating [9,10] or coherent multiple scattering is important [11–13]. In such cases, molecular dynamics (MD) simulation is often employed. However, the *ab initio* molecular dynamics (AIMD) is not feasible for simulating thermal transport due to its high computational cost considering the length and time scales of transport processes. MD simulations using EPs have been performed [14–17], but the accuracy of EPs for thermal transport in disordered phase has not been rigorously validated. A recent study shows that even atomistic structures of amorphous Si that is melt-quenched using MD with EPs and AIMD remarkably differ from each other. [18]

Machine learning potentials (MLPs) have emerged as an alternative to overcome the aforementioned problems [19,20]. They can describe multiple phases with an accuracy comparable to *ab initio* calculations, while the cost is much cheaper than the density functional theory (DFT) calculations. MLPs have been shown to successfully describe properties of materials with an accuracy of ~1 meV/atom for energy [19–33]. Notably, several previous studies showed that MLPs can have higher accuracy and reproduce correct harmonic force constants for crystalline phase [20,22,24,32]. However, to our best knowledge, the MLPs have not been examined for thermal conductivity of crystalline phase that is largely affected by subtle anharmonic interatomic interaction and thus require even higher accuracy. Also, the interatomic force constants in those previous studies were examined in single crystalline case only and the capability of MLPs in predicting interatomic interaction in multi-phases including ordered and disordered phases has not been evaluated. Here, we demonstrate that Gaussian approximation potential (GAP) [20], which is a type of MLP, has a high accuracy and flexibility to describe interatomic interactions with a DFT accuracy in both a perfect crystalline Si and crystalline Si with vacancies.

We briefly present the overall procedure of developing GAP here. Further details can be found in literature [20,22,34,35]. Our goal is to develop a machine-learning potential based on the density functional theory calculation data with minimum human intervention. The GAP method has two major components. First, a set of descriptors for atomistic structure is developed so that (1) it satisfies the translation, rotation, and permutation invariances and (2) the atomistic structures are captured precisely while the high dimensional atomic position data are mapped onto relatively low dimensional space. Second, the mapping between environments and the target quantities such as atomic energies and forces is done by a kernel which is associated with the similarity of environments. We use the smooth overlap of atomic positions (SOAP) [34]. One advantage of GAP is that the training is performed using simple linear algebra rather than iterative nonlinear optimization of a multimodal function as in the case of neural networks. The hyperparameters used for training a GAP, including those used in SOAP kernel, are listed in Table I. The meaning of all symbols can be found in Ref. [22].

Table I. Hyperparameters used in training GAP potential.

| Atomic environment kernel | SOAP |
|---|---|
| $r_{cut}$ | 6.0 Å |
| $d$ | 1.0 Å |
| $\sigma_v^{energy}$ | 0.001 eV/atom |
| $\sigma_v^{force}$ | 0.1× (force on each atom) |
| $\xi$ | 4 |
| $\delta$ | 3.0 eV |
| $\sigma_{atom}$ | 0.5 Å |
| $n_{max}$ | 12 |
| $l_{max}$ | 12 |
| Representative environments | 4000 |
| Sparse method | CUR |

The GAP was trained with the database that contains energies and forces from DFT calculation for 171 and 82 snapshots of perfect crystal and crystal with a single vacancy, respectively. Each snapshot consists of 128 atoms for the perfect crystal and 127 atoms for the crystal with a single vacancy, giving 21888 and 9984 environments, respectively. The data was taken from AIMD simulations at 300 K with a time step of 0.5 fs using VASP package [36]. To reduce correlation between snapshots, each snapshot was taken every 100 timesteps. The details of DFT calculation can be found in the Supplementary Information.

For the purpose of validating the GAP with DFT results, we performed lattice dynamics calculation using the interatomic force constants from the GAP and DFT calculation. For comparison, we also present the results from commonly used empirical potentials: the SW [3], the original version of Tersoff [4] and a modified version of the Tersoff potential [5]. The second- and third-order force constants were calculated using the finite displacement method [37]. The three-phonon scattering rate and the thermal conductivity were calculated using the ShengBTE package with a 28×28×28 mesh for sampling the first Brillouin zone [38]. The phonon-vacancy scattering rates were calculated using the exact Green's function approach [39,40]. The phonon-vacancy scattering rate is given as $\left(t_\lambda^{vac}\right)^{-1} = -\frac{f\Omega}{\omega V_d}\text{Im}\{\langle\lambda|T^+(\omega^2)|\lambda\rangle\}$, where $f$ is the volumetric fraction of vacancies, $\Omega$ is the volume of the unit cell that is used to normalize the phonon eigenstates $|\lambda\rangle$, $\omega$ is the angular frequency of phonons, $V_d$ is the volume of the vacancy. $T^+$ is a matrix defined by $T^+ = (I - VG_0^+)V$, where $I$ is the identity matrix, $V$ is the perturbation matrix, and $G_0^+$ is the retarded Green's function of the unperturbed crystal.

In Figure 1, we present the phonon dispersion from the DFT, GAP, and empirical potentials. The phonon dispersion calculated using the GAP is in excellent agreement with the DFT results, indicating that the GAP is capable of predicting very accurate second order force constants. This was also reported in the recent studies using GAP for silicon [22] and the ferromagnetic iron [32]. In addition to the second order force constants, the GAP also performs well for third-order force constants, showing the accuracy comparable to the DFT calculation, as demonstrated with the Grüneisen parameter shown in Figure 2a. The GAP can predict the Grüneisen parameter with an average relative error of 2% with respect to the DFT in the temperature range of 100-1000 K. However, the average relative error for SW, Tersoff and modified Tersoff potentials are 165%, 800% and 280%, respectively. The correct Grüneisen parameter does not necessarily lead to correct anharmonic interactions among phonon modes and thermal conductivity. Thus, we show the thermal conductivity of perfect crystalline silicon as a function of temperature in Figure 2b. In the temperature range of 100 to 1000 K, the thermal conductivity using the GAP shows very good agreement with the results from DFT and experiments by Glassbrenner and Slack [6]. To further evidence accurate third-order force constants from the GAP, the three-phonon scattering rates at 300 K are shown in Figure 3a. While the empirical potentials show poor agreement with the DFT, the three-phonon scattering rates from the GAP is almost identical with those from the DFT calculation. We also report the normalized cumulative thermal conductivity as a function of phonon mean free path (MFP) (see Figure 3b), showing that the GAP can predict accurate modal thermal conductivity not just overall thermal conductivity.

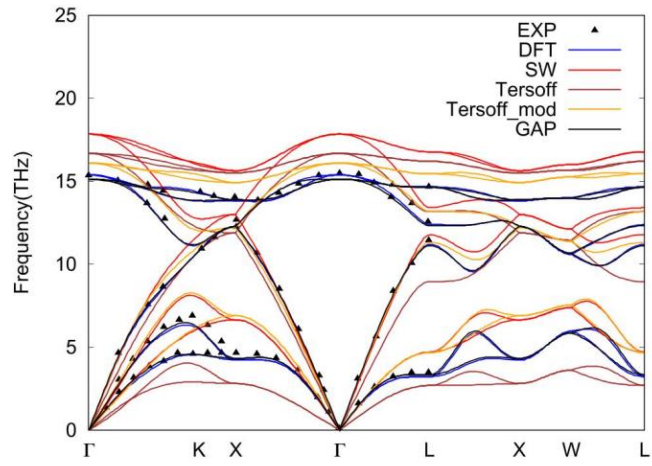

Figure 1. Phonon dispersion of crystalline silicon showing the excellent agreement between GAP and experimental data.

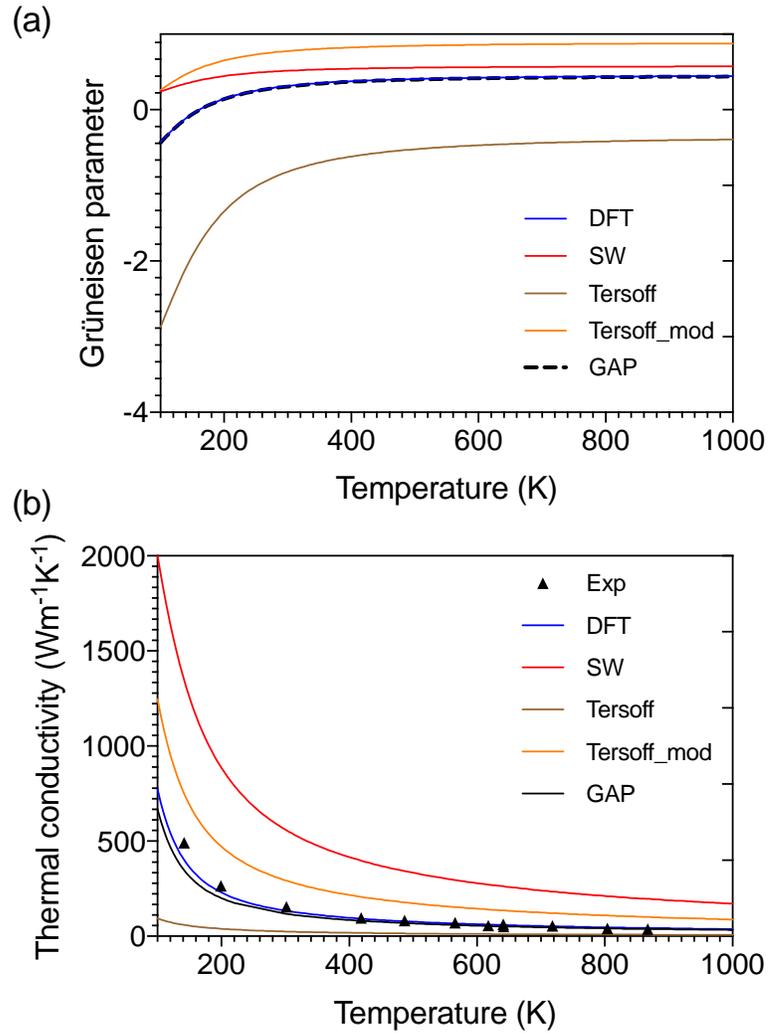

Figure 2. (a) Grüneisen parameter and (b) Thermal conductivity of crystalline silicon as a function of temperature from different potentials, DFT and experiment [6].

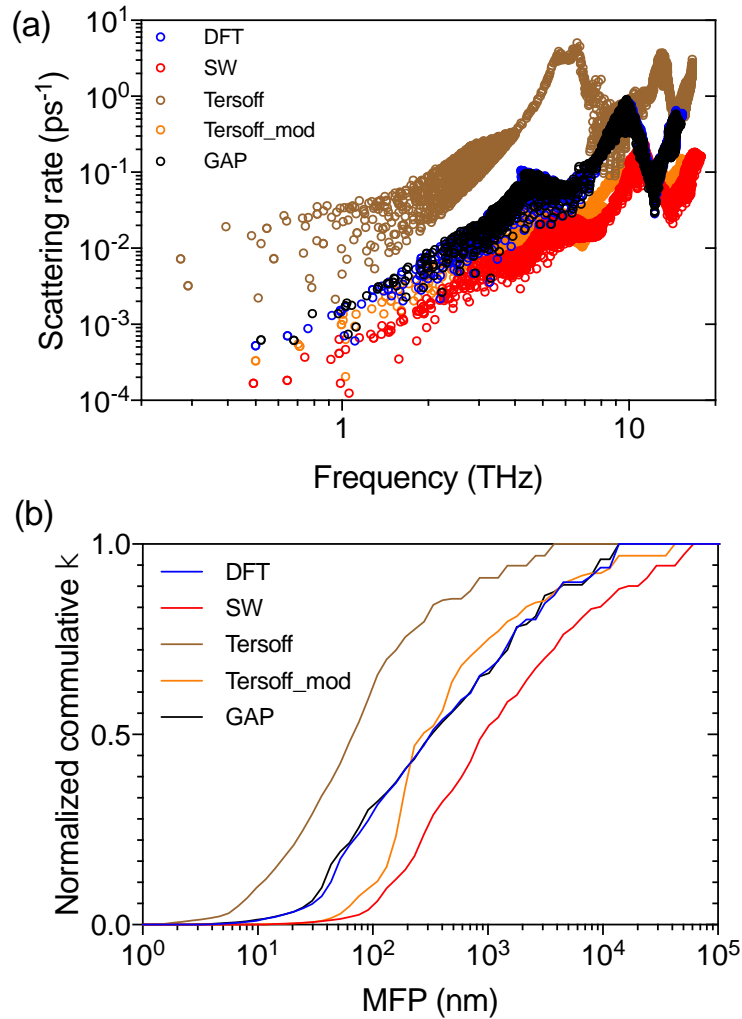

Figure 3. (a) Three-phonon scattering rates of crystalline silicon using the GAP, empirical potentials and DFT at 300 K (b) Normalized cumulative thermal conductivity as a function of phonon mean free path.

To evaluate the accuracy of GAP for crystalline silicon with single vacancy, we show the phonon-vacancy scattering rates at a vacancy concentration of 0.01% in Figure 4a. Overall, the phonon scattering rates using the GAP show a good agreement with the DFT results. Both versions of the Tersoff potential do not perform well, particularly in the low frequency region below 5 THz. The SW potential is better than the Tersoff potentials in terms of accuracy, but still shows noticeable difference from the results using GAP and DFT.

The thermal conductivity of silicon at 300 K in the presence of various volume fractions of vacancy using different potentials and DFT is shown in Figure 4b. We assume elastic scattering by a single vacancy, determined by the harmonic force constants. We also assume that the vacancies are isolated from each other, meaning that there is no overlap in the strain field induced by each vacancy and the multiple scattering by vacancies does not occur. The GAP shows an average relative error of 6% over the all vacancy concentrations, demonstrating significantly higher accuracy over empirical potentials having relative errors of 206%, 26.5% and 297% for SW, Tersoff and modified Tersoff, respectively. In Fig. 4c, we show the thermal resistivity contribution from phonon-vacancy scattering that can be extracted using the Mathiessen's rule, $k^{-1} = k^{-1}_{ph\text{-}ph} + k^{-1}_{ph\text{-}v}$ where $k^{-1}_{ph\text{-}ph}$ and $k^{-1}_{ph\text{-}v}$ represent the thermal resistivity due to three-phonon scattering and phonon-vacancy scattering, respectively. Figure 4c clearly shows that the empirical potentials fail to accurately predict the thermal resistivity due to phonon-vacancy scattering, resulting in the inaccurate thermal conductivity of defected crystalline silicon in Figure 4b. However, from Figure 2b, Figure 4b, and Figure 4c, the GAP is able to describe the thermal resistivity from both three-phonon and phonon-vacancy scattering processes. This highlights another strength of GAP in addition to its high accuracy; it is flexible enough that the single interatomic potential can accurately describe multiple atomistic configurations.

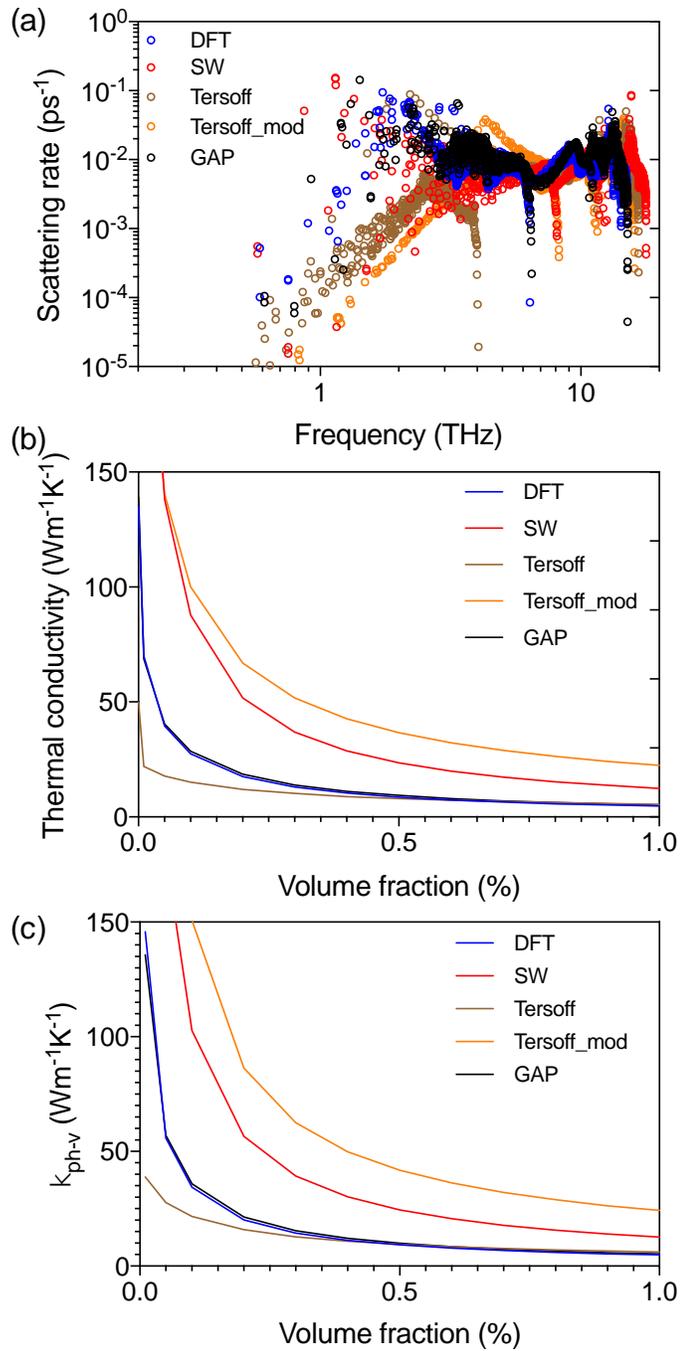

Figure 4. (a) Phonon-vacancy scattering rates at 300 K assuming the volumetric vacancy concentration of 0.01% (b) Thermal conductivity of silicon at 300 K as a function of vacancy concentration (c) Thermal conductivity at 300 K when phonon-vacancy scattering is the only scattering mechanism.

With the high accuracy and flexibility as well as the translation, rotation, and permutation invariances implemented in its descriptor, the GAP can be used to study phonon transport in a much larger system. It often requires more than thousands of atoms to represent the randomly distributed defects, and it is not feasible to study phonon transport in such a large system using DFT calculations due to extremely high computational cost. In Figure 5, we use the GAP potential to calculate the participation ratio of vibrational eigenmodes, which measures the extent of localization of vibrational eigenmodes, in a supercell of 4064 atoms (i.e., 32 randomly distributed vacancies in otherwise perfect crystalline 4096 atoms supercell). For comparison, we also use the SW potential for the calculation of participation ratio. The participation ratio from those two potentials show a remarkable difference in the frequency range from 1 to 4 THz which is known as the frequency spectrum significantly contributing to total thermal transport in crystalline Si at 300 K [8]. The much smaller participation ratio of the GAP compared to SW is consistent with the phonon-vacancy scattering rate in Figure 4(a); the phonon-vacancy scattering rate using the GAP is one order of magnitude larger than that using the SW in the frequency range of 1 to 4 THz. The direct comparison between the GAP and SW for the phonon-vacancy scattering is also shown in the Supplementary Figure S1.

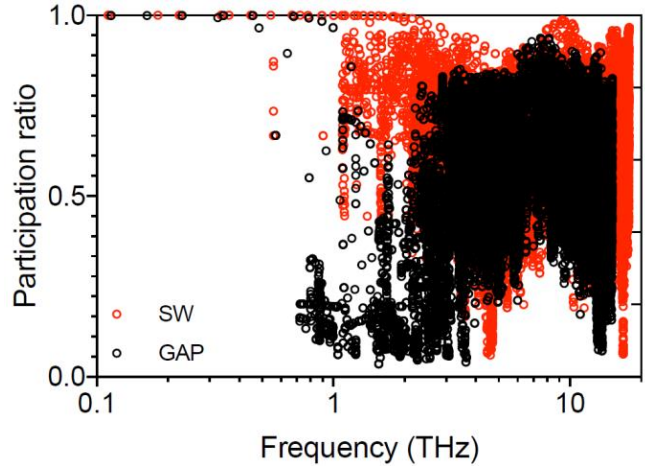

Figure 5. Participation ratio of vibrational eigenmodes in a supercell of 4064 atoms (i.e., 32 randomly distributed vacancies in otherwise perfect crystalline 4096 atoms supercell) using SW and GAP.

We also briefly compare the computational cost of the GAP and DFT. Evaluating forces in 128 atoms supercell using the GAP implementation in the LAMMPS package on one Intel Haswell (E5-2695 v3) CPU costs 3.5 sec. This is much more expensive than SW, Tersoff, and modified Tersof potentials requiring 0.0005, 0.0012, and 0.0011 sec., respectively. However, the computational cost of the GAP is 5 orders of magnitude faster than the same DFT calculation using VASP package.

In summary, we report that single interatomic potential developed using Gaussian approximation can accurately describe thermal transport in both perfect crystalline Si and crystalline Si with vacancies. We demonstrate the high accuracy of the GAP through various lattice dynamical properties including phonon dispersion, Grüneisen parameter, three-phonon scattering rates, phonon-vacancy scattering rates, and resulting thermal conductivity of Si. All of these results show an excellent agreement with the results from DFT calculations, particularly for three-phonon scattering rates which are determined by subtle anharmonic interactions of phonons and thus require high accuracy of interatomic potential. By contrast, the commonly used empirical potentials (SW, Tersoff, and modified Tersoff) show relatively poor agreement with the DFT results. In addition to the high accuracy, we note that the GAP is flexible enough to describe lattice dynamics for different atomistic configurations: perfect crystalline Si and crystalline Si with vacancies in our case. To the best of our knowledge, no potential has been developed with the DFT accuracy and flexibility for describing lattice dynamics of different atomic configurations. Our work shows that the GAP can extend the high predictive power of *ab initio* calculation for phonon transport, which has been demonstrated for perfect crystalline materials [41], to large systems that contain randomly distributed defects.

**Acknowledgment**

We acknowledge support from National Science Foundation (Award No. 1705756 and 1709307). The simulation was performed using the Linux clusters of Extreme Science and Engineering Discovery Environment (XSEDE) through allocation TG-CTS180043 and Center for Research Computing at the University of Pittsburgh. We thank Gabór Csányi for helpful discussion.